\documentclass[preprint,12pt]{elsarticle}

\usepackage{graphicx}
\usepackage{amssymb, amsthm}
\biboptions{compress}

\journal{\hspace{-3.5cm} \raisebox{-1mm}{\begin{tikzpicture} \draw [fill = white, white] (0, 0) rectangle (3.5, 0.4); \end{tikzpicture}}}

\usepackage{amsmath}
\usepackage{url}
\usepackage{color}
\usepackage{tensor}
\usepackage{mathrsfs}
\usepackage{nicefrac}
\usepackage{tikz-cd}
\usepackage{tikz}
\usetikzlibrary{decorations.markings}
\usetikzlibrary{matrix,arrows}
\usepackage{array}
\usepackage{ulem}
\usepackage{placeins}

\usepackage{pdflscape}

\usepackage[latin1]{inputenc}      
\usepackage[T1]{fontenc}    
\usepackage{ae,aecompl}                          
\usepackage{dsfont}
\usepackage{appendix}
\usepackage{amsfonts,amsmath,latexsym,bbm}
\usepackage{amssymb}
\usepackage{accents}
\usepackage{graphics}
\usepackage{graphicx}
\usepackage{booktabs,longtable}
\usepackage{dcolumn}
\usepackage{enumitem}
\usepackage{upgreek}
\usepackage{tensor}

\usepackage{tikz}
\usepackage{tikz-cd}
\usepackage{tikz-3dplot}
\usetikzlibrary{decorations.markings}
\usetikzlibrary{matrix,arrows}
\usetikzlibrary{calc}
\usetikzlibrary{shapes}
\usetikzlibrary{positioning}
\usetikzlibrary{fit}

\usepackage{pgfplots}
\pgfplotsset{width=7cm,compat=1.5.1}

\usetikzlibrary{decorations.pathreplacing}
\makeatletter
\def\underbrace#1{%
  \@ifnextchar_{\tikz@@underbrace{#1}}{\tikz@@underbrace{#1}_{}}}
\def\tikz@@underbrace#1_#2{%
  \tikz[baseline=(a.base)] {\node[inner sep=2] (a) {\(#1\)};
  \draw[thick,line cap=round,decorate,decoration={brace,amplitude=4pt}]
    (a.south east) -- node[pos=0.5,below,inner sep=7pt] {\(\scriptstyle #2\)} (a.south west);}}
\makeatother

\usepackage[linktocpage=true]{hyperref}

\makeatletter
\renewcommand*{\eqref}[1]{%
  \hyperref[{#1}]{\textup{\tagform@{\ref*{#1}}}}%
}
\makeatother

\setlist{font=\normalfont\itshape} 

\renewcommand{\vec}[1]{\boldsymbol{#1}}
\newcommand{\tsr}[1]{\overset\leftrightarrow{#1}}
\newcommand{\ext}{_{\rm ext}}
\newcommand{\tot}{_{\rm tot}}
\newcommand{\ind}{_{\rm ind}}

\newcommand{\h}{\hspace{1pt}}
\newcommand{\hh}{\hspace{0.5pt}}
\newcommand{\mh}{\hspace{-1pt}}

\newcommand{\de}{\mathrm d}

\newcommand{\lar}[1]{\textnormal{\mbox{\large $#1$}}}

\definecolor{oldgray}{gray}{0.4}

\newcommand{\T}{_{\mathrm T}}
\renewcommand{\L}{_{\mathrm L}}

\DeclareMathAlphabet{\mathbbmsl}{U}{bbm}{m}{sl}

\numberwithin{equation}{section}

\begin{document}

\begin{frontmatter}

%% Title, authors and addresses

%% use the tnoteref command within \title for footnotes;
%% use the tnotetext command for the associated footnote;
%% use the fnref command within \author or \address for footnotes;
%% use the fntext command for the associated footnote;
%% use the corref command within \author for corresponding author footnotes;
%% use the cortext command for the associated footnote;
%% use the ead command for the email address,
%% and the form \ead[url] for the home page:
%%
%% \title{Title\tnoteref{label1}}
%% \tnotetext[label1]{}
%% \author{Name\corref{cor1}\fnref{label2}}
%% \ead{email address}
%% \ead[url]{home page}
%% \fntext[label2]{}
%% \cortext[cor1]{}
%% \address{Address\fnref{label3}}
%% \fntext[label3]{}

\title{Covariant Response Theory and the \\[1pt]
Boost Transform of the Dielectric Tensor}

%% use optional labels to link authors explicitly to addresses:
%% \author[label1,label2]{<author name>}
%% \address[label1]{<address>}
%% \address[label2]{<address>}

\author[freiberg]{R.~Starke}
\ead{Ronald.Starke@physik.tu-freiberg.de}

\author[heidelberg]{G.A.H.~Schober\corref{cor1}}
\ead{schober@physik.rwth-aachen.de}

\cortext[cor1]{Corresponding author.}

\address[freiberg]{Institute for Theoretical Physics, TU Bergakademie Freiberg, Leipziger Stra\ss e 23, \\ 09599 Freiberg, Germany}
\address[heidelberg]{Institute for Theoretical Physics, Heidelberg University, Philosophenweg 19, \\ 69120 Heidelberg, Germany \vspace{-0.6cm}}

\begin{abstract}
After a short critique of the Minkowski formulae for the electromagnetic constitutive laws in moving media, we argue that in actual fact the problem of Lorentz-covariant electromagnetic response theory is automatically solved within the framework of modern microscopic  electrodynamics of materials. As an illustration, we first rederive the well-known relativistic transformation behavior of the microscopic conductivity tensor. Thereafter, we deduce from first principles the transformation law of the wavevector- and frequency-dependent dielectric {\it tensor} under Lorentz boost \mbox{transformations.}
\end{abstract}

\begin{keyword}
%% keywords here, in the form: keyword \sep keyword
electrodynamics of media \sep special relativity

%% MSC codes here, in the form: \MSC code \sep code
%% or \MSC[2008] code \sep code (2000 is the default)

\end{keyword}

\end{frontmatter}

%%
%% Start line numbering here if you want
%%
% \linenumbers

%% main text

\newpage
\tableofcontents

\newpage
\section{Introduction}

In contrast to condensed matter physics, special relativity is nowadays usually associated with phenomena involving high energies or momenta.
Surprisingly, however, it was only three years after the introduction of special relativity by A.~Einstein in 1905 \cite{Einstein1905} that
the relativistic generalization of electromagnetic constitutive relations came under scrutiny. In fact, already in 1908, H.~Minkowski derived
the seemingly relativistic material relations which now bear his name \cite{Minkowski}. Since then, this topic continues to hold its importance:
On the one hand, electromagnetic material properties in the relativistic r\'{e}gime are most naturally relevant to {\itshape plasma physics} (see e.g.~\cite{Melrose1Book,Melrose2Book}).
On the other hand, it is well known that {\itshape moving media}  lead to qualitatively new effects which have observable consequences even at ``non-relativistic'' velocities. This fact has been known
at least since the famous Fizeau experiment in 1851 \cite{Fizeau} and its subsequent interpretation in the light of special relativity
by M.~von Laue in 1907 \cite{Laue}. At present, the ``relativistic response'' of electrons is studied in various materials such as graphene \cite{Dani} or Rashba semiconductors \cite{Lee,Schwalbe}. Correspondingly,
relativistic {\itshape electronic structure theory} in general \cite{Schwerdtfeger1,Schwerdtfeger2} and relativistic {\itshape density functional theory} in particular \cite{Engel}
are by now well-established branches of first-principles materials science.

On the basis of all these developments, it has become highly desirable to develop a fully relativistic electromagnetic response theory.
Therefore, the Minkowski formulae developed long {\it before} the advent of modern ab initio electronic structure theory 
(see e.g.~\cite{Hafner08,Hafner10,Kohanoff,Martin} for introductions) automatically come back into the focus. In this context, 
it is noteworthy that the ab initio calculation of electromagnetic response functions
(for classical articles see \cite{Adler,Wiser,Hanke,Strinati}, and for modern textbooks see \cite{Bechstedt,Giuliani,Bruus})
relies on a {\it microscopic approach to electrodynamics of media} \cite{ED1,ED2}, according to which linear electromagnetic response properties are not given by shear constants
(as it had traditionally been assumed) but by {\it response functions} constituting entire {\it non-local integral kernels}. This fact---which renders the historical derivation of the Minkowski formulae inapplicable---completely eluded
researchers at the time when these had originally been developed, and hence the ensuing relativistic constitutive equations require a thorough re-investigation from the modern microscopic point of view. This is precisely the principal objective~of the present article.

Concretely, we will begin this article in Sct.~\ref{Sec_Standard} with a short explanation of the Minkowski formulae for the relativistic response,
and then elaborate on our criticism of them. In the subsequent Sct.~\ref{Sec_Fundamentals}, we will briefly introduce microscopic electromagnetic linear response theory
in terms of the {\it fundamental response tensor} in a manifestly covariant manner. With these preparations, we will in Sct.~\ref{Sec_CovRT} 
systematically develop the {\itshape covariant Cartesian response theory} by deriving the relativistic tansformation properties
of the conductivity tensor and the dielectric tensor. The first Appendix~\ref{app_srt} summarizes the essential formulae of special relativity which are needed in this article.
Finally, as an illustration of the microscopic theory developed in this article, Appendix~\ref{sec_fizeau} contains a straightforward rederivation of the famous Fizeau result for the refractive index of moving media.

\section{Standard Approach}\label{Sec_Standard}
\subsection{Minkowski formulae}

In deriving the Minkowski formulae, one starts from the usual electric and magnetic {\itshape material relations} (or {\itshape constitutive relations}) written in the form
\begin{align}
\vec D'&=\varepsilon_0 \h \varepsilon_{\rm r}\h \vec E'\,,\label{eq_naive1}\\[3pt]
\vec B'&=\mu_0 \h \mu_{\rm r}\h \vec H'\,,\label{eq_naive2}
\end{align}
as well as Ohm's law in the form
\begin{equation}
\vec j'=\sigma\h\vec E'\,.\label{eq_naive3}
\end{equation}
These relations are assumed to hold in a certain inertial frame, which we identify with the ``primed'' reference frame. The quantities $\vec D$, $\vec H$ and $\vec j$ denote the displacement field, the magnetic field and the electric current density, while $\vec E$ and $\vec B$ are the electric field and the magnetic induction. Furthermore, the permittivity $\varepsilon_{\rm r}$\hh, the permeability $\mu_{\rm r}$\hh, and the conductivity $\sigma$ are assumed to be {\itshape constants} in the above formulae. The problem now consists in deriving analogous material relations in another inertial frame, the ``unprimed'' frame, 
which moves relatively to the primed frame with the velocity $-\vec v$ (see Appendix \ref{app_boost}). For this purpose, one assumes that $\vec D/\varepsilon_0$ and $\mu_0\hh\vec H$ transform precisely in the same way as $\vec E$ and $\vec B$, respectively (see Eqs.~\eqref{eq_trafo1} and \eqref{eq_trafo2} derived in Appendix \ref{app_emtrafo}). Hence,
\begin{equation} \label{put_1}
\vec D'=\gamma\left(\vec D+\frac{\vec v\times\vec H}{c^2}\right)-(\gamma - 1) \, \frac{\vec v \h (\vec v\cdot\vec D)}{|\vec v|^2}\,,
\end{equation}
and
\begin{equation} \label{put_2}
\vec H'=\gamma\h\hh(\vec H-\vec v\times\vec D)- (\gamma - 1) \, \frac{\vec v \h (\vec v \cdot \vec H)}{|\vec v|^2} \,.
\end{equation}
We note in passing that this assumption is completely in accord with the {\it Fundamental Field Identifications} introduced by the Functional Approach to electrodynamics of media \cite{ED1,ED2,Refr,EffWW,EDOhm}. By putting these transformation formulae into Eqs.~\eqref{eq_naive1} and \eqref{eq_naive2}, one arrives at the well-known relations
\begin{align}
\vec D + \frac{\vec v\times\vec H}{c^2} &=\varepsilon_0 \h \varepsilon_{\rm r} \h (\vec E +\vec v\times\vec B)\,, \label{eq_Minkowski1}\\[5pt]
\vec B - \frac{\vec v\times \vec E}{c^2} &=\mu_0 \h \mu_{\rm r} \h (\vec H - \vec v\times\vec D)\,. \label{eq_Minkowski2}
\end{align}
Here, we have used the equalities
\begin{equation} \label{eq_breakdown_1}
 \vec v\cdot\vec D \h = \h \vec v \cdot \vec D' \h = \h \varepsilon_0 \h \varepsilon_{\rm r}\h (\vec v\cdot\vec E') \h = \h \varepsilon_0 \h \varepsilon_{\rm r}\h (\vec v\cdot\vec E) \,,
\end{equation}
as well as \smallskip
\begin{equation}
\vec v\cdot\vec B = \mu_0 \h \mu_{\rm r} \h (\vec v \cdot \vec H)  \label{eq_breakdown_2} \,, \smallskip
\end{equation}
which follow from the fact that the electric and magnetic field components parallel to the boost velocity $\vec v$ stay constant under the Lorentz transformation (see Eqs.~\eqref{eq_boost_parr_1}--\eqref{eq_boost_parr_2}).
By similar considerations, one shows that
\begin{align}
\vec j-\rho \h \vec v&=\gamma \h \sigma\left(\vec E + \vec v\times\vec B-\frac{\vec v  \h (\vec v\cdot\vec E)}{c^2}\right), \label{eq_OhmGen}
\end{align}
where $\rho$ denotes the charge density. The above Eqs.~\eqref{eq_Minkowski1}--\eqref{eq_Minkowski2} are known as {\itshape Minkowski formulae}, 
and they are well established in the textbook literature,
see e.g.~\cite[p.~234, Eqs.~(9.7)--(9.8)]{Bredov}, \cite[Problem 12.68]{Griffiths}),
\cite[Eqs.~(5.127)--(5.128)]{Toptygin}, \cite[\S\h22.8.1]{Zangwill}, and 
\cite[Eqs.~(12.6)]{Stephani} (for an alternative version compare also \cite[Eqs.~(12.7)]{Stephani} with \cite[Eqs.~(5.25)--(5.28)]{RebhanRel}, \cite[Eqs.~(2.138)]{Roemer}). By contrast,
Eq.~\eqref{eq_OhmGen} is the so-called {\it generalized Ohm law}, see e.g.~\cite[Eq. (5.28)]{RebhanRel}, \cite[Eq.~(2.132)]{Roemer}, \cite[Problem 11.16]{Jackson}, and \cite[Problem 9-15]{Tsang}.
In particular, the Minkowski formulae are supposed to yield ``fully relativistic constitutive relations'' \cite[p.~859]{Zangwill}.
Before coming to our account of relativistic electromagnetic response theory, we will in the next subsection develop our criticism of these formulae. 

\subsection{Problems}

Our criticism of the Minkowski formulae is based on four main arguments to be spelled out now:

\begin{enumerate}[listparindent=\parindent,parsep=0pt,itemsep=1em]
 \item[(i)]{\itshape Tensorial response functions.}
The derivation of the Minkowski formulae breaks down for tensorial response functions, because in this 
case the field components parallel and orthogonal to the boost velocity $\vec v$ do not decouple. For example, by generalizing Eq.~\eqref{eq_naive1} to
\begin{equation} \label{more_gen}
 \vec D' =\varepsilon_0 \h\hh \tsr{\varepsilon_{\rm r}}\h \vec E'
\end{equation}
with a $(3 \times 3)$-matrix $\tsr{\varepsilon_{\rm r}}$\hh, we find that in general,
\begin{equation}
 \vec v \h (\vec v \cdot \vec D') \h  
 = \h \varepsilon_0 \h\hh \vec v \h (\vec v \cdot (\hh\tsr{\varepsilon_{\rm r}} \h \vec E')) \h \not 
 = \h \varepsilon_0 \h\hh \tsr{\varepsilon_{\rm r}} \h\hh \vec v \h (\vec v \cdot \vec E') \,,
\end{equation}
and hence Eqs.~\eqref{eq_breakdown_1} and \eqref{eq_breakdown_2} cannot be employed.

\item[(ii)]{\itshape Wavevector and frequency dependence.}
 In the Minkowski formulae, the response functions are treated as {\itshape constants} (i.e.~numbers), while in reality they constitute {\itshape response functions.} These are non-local integral kernels in real space,
or (for homogeneous systems) functions of the wavevector and the frequency in Fourier space. In other words, Eq.~\eqref{more_gen} has to be generalized further to
\begin{equation}
 \vec D'(\vec x, t) = \varepsilon_0 \int \! \de^3 \vec x' \! \int \! c \, \de t' \, \tsr{\varepsilon_{\rm r}}(\vec x - \vec x', \h t - t') \h \vec E'(\vec x', t') \,,
\end{equation}
or in Fourier space to
\begin{equation}
  \vec D'(\vec k, \omega) =\varepsilon_0 \h\hh \tsr{\varepsilon_{\rm r}}(\vec k, \omega) \h \vec E'(\vec k, \omega) \,.
\end{equation}
This shows that the wavevector- and frequency-dependent response functions also have to be transformed under Lorentz boosts, a
fact which is completely ignored in the Minkowski formulae.

\item[(iii)]{\itshape Inequivalent inertial frames.}
Contrary to the standard textbook lore, the Minkowski formulae actually do not constitute covariant response laws.
In fact, a (Lorentz-)covariant law is an equation which has the same form in every (inertial) frame. This, however, does not apply to the Minkowski formulae.
Quite to the contrary, via the boost parameter~$\vec v$ the Minkowski formulae single out a {\itshape preferred frame} where the na\"{i}ve response laws in the form of Eqs.~\eqref{eq_naive1}--\eqref{eq_naive3} hold.
Given an {\itshape arbitrary} inertial frame, one is thus not able to determine the response relations of the electromagnetic fields associated with that very frame. Instead, in the first place
one would have to determine the velocity $\vec v$ of the given inertial frame with respect to the preferred frame where Eqs.~\eqref{eq_naive1}--\eqref{eq_naive3} hold. In particular, this shows that in a
covariant response law, the boost parameter $\vec v$ must not enter as a matter of principle. Instead, this parameter should enter the transformation rules of all quantities involved in the response~law.

\item[(iv)]{\itshape Undefined rest frame.}
Finally, it is not clear in which frame the na\"{i}ve formulae \eqref{eq_naive1}--\eqref{eq_naive3} actually hold, i.e., what defines the preferred frame mentioned under~{\itshape (iii).} Associating this preferred frame with the ``rest frame'' may be intuitive
for the simple case of moving {\it bodies}. By contrast, in the case of an {\itshape electron gas} 
(considered as a subsystem of a crystalline system composed of electrons and nuclei) or for {\itshape relativistic plasmas,} to cite only two examples, it is not so obvious how to proceed.
In general, the response laws given by Eqs.~\eqref{eq_naive1}--\eqref{eq_naive3} are relations between {\it fields,} and fields have no ``rest frame''.
\end{enumerate}

\medskip \noindent
We remark that in previous works \cite{Gordon23, Schmutzer56, Schmutzer68, Obukhov08, Post62, Hehl1, Thompson10, Thompson11}, the problem {\itshape (i)} of the {\itshape original} Minkowski formulae \eqref{eq_Minkowski1}--\eqref{eq_Minkowski2} has been cured by switching over to a formulation in terms of manifestly Lorentz-covariant response relations (see also Refs. \cite{Antoci97, Antoci98} which take into account frequencies and wavevectors). In particular, one may consider instead of the standard relations \eqref{eq_naive1}--\eqref{eq_naive2} a generalized constitutive relation (see Refs.~\cite{Hehl1, Melrose, Fechner, Hehltensor}), which reads in an adapted notation \smallskip
\begin{equation}
 F^{\mu\nu}_{\rm ext} = \chi\indices{^{\mu\nu}_{\alpha\beta}} \h F^{\alpha\beta}_{\rm tot} \,. \smallskip
\end{equation}
Here, $F^{\mu\nu}$ denotes the field strength tensor, which incorporates both electric and magnetic field components, and $\chi\indices{^{\mu\nu}_{\alpha\beta}}$ is called the field-strength response tensor (see Sct.~\ref{sec_fsrt}).
In this article, we aim instead at a Lorentz-covariant formulation of electromagnetic response theory in terms of the original three-dimensional (Cartesian) response tensors such as the conductivity tensor or the dielectric tensor. Furthermore, our approach also takes into account the fact that these response functions are not independent of each other but related by Universal Response Relations (see Ref.~\cite{ED1} and Sct.~\ref{sec_fsrt} below). Such a fully relativistic electromagnetic response theory in terms of Cartesian response tensors has yet to be developed, a task which will be accomplished in the following.

\section{Functional Approach}\label{Sec_Fundamentals}
\subsection{Fundamental response tensor} \label{sec_fundresp}

Modern microscopic electromagnetic response theory is based on the functional dependence $j^\mu[A^\nu]$ 
of the induced four-currentr, $j^\mu\equiv j^\mu\ind$\hh, on the externally applied four-potential, $A^\nu\equiv A^\nu\ext$ 
(see e.g.~\cite{Bechstedt,Giuliani,Bruus,Melrose,Melrose73,SchafWegener}). 
This is precisely the starting point of the {\it Functional Approach} to electrodynamics of media \cite{ED1,ED2,Refr,EffWW,EDOhm},
which in this respect axiomatizes the common practice in ab initio materials physics.
Correspondingly, up to linear order in the external perturbation, we have the expansion
\begin{equation}
j^\mu(x)=\int\!\de^4 x'\,\chi\indices{^\mu_\nu}(x,x') \h A^\nu(x')\,, \label{eq_GenRespLaw}
\end{equation}
where $x = (c\hh t, \h \vec x)^{\rm T}$, and where $\chi\indices{^\mu_\nu}$ denotes the {\it fundamental response tensor} \cite{Melrose1Book,Adler,Strinati,Melrose73,Altland}. The latter can also be characterized  as the functional derivative (see Ref.~\cite[\S\h5.1]{ED1})
\begin{equation}
\chi\indices{^\mu_\nu}(x, x') = \frac{\delta j^\mu(x)}{\delta A^\nu(x')} \,.
\end{equation}
On account of the continuity equation for the induced current density and its invariance
with respect to gauge transformations of the external four-potential, the fundamental response tensor has the following general form in Fourier space \cite{Melrose1Book,ED1,Melrose73}:
\begin{equation}\label{generalform}
\chi^\mu_{~\nu}(k,k')=
\left( \!
\begin{array}{rr} 
-\, \lar{\frac{c\vec k^{\rm T}}{\omega}} \, \tsr \chi(k,k') \, \lar{\frac{c\vec k'}{\omega'}} & \lar{\frac{c\vec k^{\rm T}}{\omega}} \,  \tsr \chi(k,k') \, \\[10pt] 
- \, \tsr \chi(k,k') \, \lar{\frac{c\vec k'}{\omega'}} & \, \tsr \chi(k,k') \, 
\end{array} \right),
\end{equation}
where $k = (\omega/c, \h \vec k)^{\rm T}$. Therefore, the fundamental response tensor is completely determined by its spatial part, the {\it current response tensor} given by
\begin{equation}
\tsr\chi(k,k')=\frac{\delta\vec j(k)}{\delta\vec A(k')}\,.
\end{equation}
Furthermore, as the external four-potential contains the complete information about the externally
applied electromagnetic perturbation, Eq.~\eqref{eq_GenRespLaw} constitutes the most general linear electromagnetic response law, which includes all effects of inhomogeneity,
anisotropy, magnetoelectric cross-coupling and relativistic retardation. As shown explicitly in Ref.~\cite{ED1}, all linear electromagnetic response functions can be calculated analytically from the fundamental response tensor
by means of {\itshape universal} (i.e., frame-, model- and material-independent) {\itshape response relations.}

At the same time, Eq.~\eqref{eq_GenRespLaw} is manifestly Lorentz covariant and thus holds in every inertial frame. Here it is understood that the fundamental response
tensor itself obeys a transformation law. The latter can be deduced directly from the well-known transformation law for four-vector fields,
\begin{align}
j^{\mu'}(x') &=\Lambda\indices{^{\mu'}_{\mu}} \, j^\mu(x)\,, \label{usual_1} \\[5pt]
A^{\mu'} (x')&=\Lambda\indices{^{\mu'}_{\mu}} \h A^\mu(x)\,, \label{usual_2}
\end{align}
where $\Lambda$ is the Lorentz transformation under consideration, and $x' = \Lambda \h x$. It turns out that the transformation law of the fundamental response tensor is of the tensorial type and reads explicitly
\begin{equation}
 \chi\indices{^{\mu'}_{\nu'}}(x',y') = \Lambda\indices{^{\mu'}_{\mu}} \, \Lambda\indices{_{\nu'}^{\nu}} \h \chi\indices{^{\mu}_{\nu}}(\Lambda^{-1} x', \h \Lambda^{-1} y') \,.
\end{equation}
With this transformation law and with Eqs.~\eqref{usual_1}--\eqref{usual_2}, the fundamental response law given by Eq.~\eqref{eq_GenRespLaw} holds in the primed coordinate system as well. 
In principle, these straightforward considerations already solve the problem of formulating a covariant linear electromagnetic response theory on the most fundamental level.

\subsection{Field strength response tensor} \label{sec_fsrt}

Traditionally, however, the usage of the fundamental response tensor which determines the induced four-current in terms of the external four-potential is not very widespread. Instead, one usually works with response functions
of induced electric and magnetic fields with respect to external electric and magnetic fields. A natural way to incorporate
these into our formalism lies in the transition to the manifeslty covariant {\it field strength response tensor} \cite{Melrose1Book,ED1}, \smallskip
\begin{equation}
\chi\indices{^{\mu\nu}_{\alpha\beta}}(x,x')
=\frac{\delta F^{\mu\nu}\ind(x)}{\delta F^{\alpha\beta}\ext(x')} \,, \smallskip  \label{eq_fsrt}
\end{equation}
where $F^{\mu\nu}$ denotes the (induced or external) field strength tensor (see Appendix \ref{app_emtrafo}). The corresponding linear expansion  reads
\begin{equation}
F_{\rm ind}^{\mu\nu}(x) = \int\!\de^4 x'\,\chi\indices{^{\mu\nu}_{\alpha\beta}}(x, x') \h F_{\rm ext}^{\alpha\beta}(x') \,.
\end{equation}
It is straightforward to show that the field strength response tensor
can be expressed analytically in terms of the fundamental response tensor. The concrete expressions can be written in real space as \cite[Eq.~(5.39)]{ED1}
\begin{equation}
\begin{aligned}
\chi\indices{^{\mu\nu}_{\alpha\beta}}(x, x') & =
\frac{1}{2\mu_0} \int \! \de^4 y \int \! \de^4 y' \, \mathbbmsl D_0(x-y) \\[3pt]
& \quad \, \times \Big(
\partial^\mu \h \partial'_\alpha \h \chi\indices{^\nu_\beta}(y,y')-
\partial^\nu \h \partial'_\alpha \h \chi\indices{^\mu_\beta}(y,y') \\[1pt]
& \quad \quad \quad - \partial^\mu \h \partial'_\beta \h \chi\indices{^\nu_\alpha}(y,y')+
\partial^\nu \h \partial'_\beta \h \chi\indices{^\mu_\alpha}(y,y')
\Big) \, \mathbbmsl D_0(y'-x') \,,
\end{aligned}
\end{equation}
and in Fourier space as (see \cite[Eq.~(1.5.16)]{Melrose1Book} or \cite[Eq.~(5.40)]{ED1})
\begin{align} \label{eq_chichi}
\chi\indices{^{\mu\nu}_{\alpha\beta}}(k, k') & = -\frac{1}{2\mu_0} \, \mathbbmsl D_0(k) \, \Big(
k^\mu \h \chi\indices{^\nu_\alpha}(k,k') \h k'_\beta -
k^\nu \h \chi\indices{^\mu_\alpha}(k,k') \h k'_\beta \\ \nonumber
& \hspace{3cm} \, - k^\mu \h \chi\indices{^\nu_\beta}(k,k') \h k'_\alpha +
k^\nu \h \chi\indices{^\mu_\beta}(k,k') \h k'_\alpha
\Big) \, \mathbbmsl D_0(k') \,.
\end{align}
Here, $\mathbbmsl D_0$ denotes the Green function of the d'Alembert operator, which is given explicitly in momentum space by \cite[\S\h3.1]{ED1}
\begin{equation}
\mathbbmsl D_0(\vec k, \omega) = \frac{1}{\varepsilon_0} \h \frac{1}{-\omega^2 + c^2 |\vec k|^2} \,.
\end{equation}
By invoking the general form \eqref{generalform} of the fundamental response tensor and reading out the electric and magnetic fields separately, 
we thus obtain the following {\itshape partial} response functions (cf.~\S\h\ref{subsec_physresp}):
 \begin{align}
\frac{\delta E^i_{\rm ind}(\vec k, \omega)}{\delta E^j_{\rm ext}(\vec k'\mh, \omega')} & \h = \h 
  \frac{\omega^2 \delta_{im} - c^2 k_i k_m}{\omega^2 - c^2 |\vec k|^2} \ 
 \frac{\chi_{mn}(\vec k,\omega; \h \vec k'\mh,\omega')}{-\varepsilon_0 \h\hh \omega \h \omega'} \
 \frac{\omega'^2 \delta_{nj} - c^2 k'_n k'_j}{\omega'^2 - c^2 |\vec k'|^2} \,, \label{eq_part_der1} \\[5pt]
\frac{1}{c} \, \frac{\delta E^i_{\rm ind}(\vec k, \omega)}{\delta B^j_{\rm ext}(\vec k'\mh, \omega')} & \h = \h 
  \frac{\omega^2 \delta_{im} - c^2 k_i k_m}{\omega^2 - c^2 |\vec k|^2} \
 \frac{\chi_{mn}(\vec k,\omega; \h \vec k'\mh, \omega') }{-\varepsilon_0 \h\hh \omega\h \omega'}\
 \frac{\epsilon_{n\ell j} \, \omega' \h c \hh k'_\ell}{\omega'^2 - c^2 |\vec k'|^2} \,, \label{eq_part_der2} \\[5pt]
c \, \frac{\delta B^i_{\rm ind}(\vec k, \omega)}{\delta E^j_{\rm ext}(\vec k'\mh, \omega')} & \h = \h 
 \frac{\epsilon_{ikm} \, \omega \h c \hh k_k}{\omega^2 - c^2 |\vec k|^2} \
 \frac{\chi_{mn}(\vec k,\omega; \h \vec k'\mh, \omega')}{-\varepsilon_0 \h\hh \omega\h \omega'} \
 \frac{\omega'^2 \delta_{nj} - c^2 k'_n k'_j}{\omega'^2 - c^2 |\vec k'|^2} \,, \label{eq_part_der3} \\[5pt]
\frac{\delta B^i_{\rm ind}(\vec k, \omega)}{\delta B^j_{\rm ext}(\vec k'\mh, \omega')} & \h = \h 
  \frac{\epsilon_{ikm} \, \omega \h c \hh k_k}{\omega^2 - c^2 |\vec k|^2} \
 \frac{\chi_{mn}(\vec k,\omega; \h \vec k'\mh, \omega') }{-\varepsilon_0 \h\hh \omega \h \omega'} \
 \frac{\epsilon_{n\ell j} \, \omega' \h c \hh k'_\ell}{\omega'^2 - c^2 |\vec k'|^2} \,, \label{eq_part_der4}
\end{align}
where we sum over all doubly appearing {\it Cartesian} indices ($i=1,2,3$).
These formulae slightly generalize the ones given in Ref.~\cite[\S\h6.5]{ED1} in that they apply also to temporally inhomogeneous systems. They are valid in every inertial frame,
and thus they constitute the searched-for relativistically covariant formulation of electromagnetic linear response theory. For the corresponding expansion of the induced electric and magnetic fields in terms of the external perturbation, we refer the interested reader to Ref.~\cite[\S\h6.6]{ED1}.

\subsection{Homogeneous limit and covariance}

As mentioned before, the microscopic electromagnetic response theory which is based on the fundamental response tensor is generally valid and in particular applies even to inhomogeneous systems. Here, ``system'' may refer to any concrete material sample with a certain geometry, shape, boundaries, etc. Correspondingly, 
the response functions involved in this general formalism depend on two wavevectors $\vec k, \h \vec k'$ and on two frequencies $\omega, \h \omega'$. In fact, even for the hypothetical case 
of a material filling out all of space (i.e., a system with no boundaries),  due to the atomic structure of the material the microscopic response functions do not necessarily have to be {\it spatially homogeneous}.
By contrast, {\itshape material properties} are always assumed to be {\it homogeneous with respect to time}. Hence, the corresponding response 
functions depend only on the difference ($t-t'$) of the two time arguments, or in the Fourier domain on only one frequency.

Unfortunately, however, this condition is not Lorentz covariant. This means, a response function which in a certain 
inertial frame depends on two wavevectors $\vec k, \h \vec k'$ but only one frequency $\omega$, does not necessarily 
have the same form in another inertial frame, where it will in general depend nontrivially on two entire four-vectors $k$ and $k'$. This problem does not arise in the {\itshape homogeneous limit,} where the response functions essentially depend on only one four-vector $k$ (see \cite[\S\h2.1]{ED1}).
Happily, we note that homogeneity is assumed anyway for most practical applications and, in particular, for the derivation 
of wave equations in media (see the discussion in Ref.~\cite[\mbox{\S\h3.2.2}]{ED2}). Without spatial and temporal homogeneity though, things become more complicated.
Then, the dependence of a response function on two frequencies implies a nontrivial dependence on two times $t$ and $t'$, as opposed to a mere dependence on the time difference ($t-t'$). Physically, this implies for example that the ``delay'' between a perturbation
and a response depends on when the perturbation is switched on! Clearly, such a response does not qualify as a material property.
In fact, it is at odds with the intuitive idea that a material property does not change in time---otherwise it would not be a material property
but a property of the concrete system \mbox{under consideration.}

We therefore come to the following conclusion:
{\it For genuinely inhomogeneous materials, there cannot be a Lorentz covariant description.}
Instead, there is a {\it preferred frame of reference} in which the spatially inhomogeneous system displays a temporal homogeneity. We stress, however, that this does not mean that the fundamental linear response law \eqref{eq_GenRespLaw} would loose its covariance or even its validity for inhomogeneous systems.
Quite to the contrary, the response law remains valid, but the corresponding response functions can then not be interpreted as material properties anymore.
Instead, they then characterize the response of the concrete inhomogeneous system under consideration.

\subsection{Physical response functions} \label{subsec_physresp}

As has been exlained in detail in Ref.~\cite[\S\h6.1]{ED1}, the {\itshape partial} response functions introduced by Eq.~\eqref{eq_fsrt} in the preceding subsection
do not correspond to {\it physical} response functions. The reason for this is that they are defined as {\it partial} derivatives, in which per definitionem the external electric and magnetic fields are varied {\it independently} of each other. In reality, however, this is not possible
because the external electromagnetic fields have to fulfill Maxwell's equations, and thus the transverse electric field
is linked to the magnetic field via Faraday's law. Consequently, although the partial response functions \eqref{eq_part_der1}--\eqref{eq_part_der4}
give rise to a valid linear expansion of the induced fields in terms of the external fields, the involved response functions do not correspond
to quantities which can be measured {\it separately}.

For this reason, {\itshape physical} response functions are necessarily given by {\it total} functional derivatives.
They serve the purpose of expressing the induced fields in terms of those parts
of the perturbation which can be varied independently (i.e., the longitudinal electric field and the magnetic field, or alternatively,
the electric field and the static part of the magnetic field, see Ref.~\cite[\S\h4.3 and \S\h6.6]{ED1}). Both practically and conceptually, physical response functions are therefore best suited for  
the description of material properties. Concretely, this means that in a linear expansion like
\begin{equation}
\vec E\ext = \tsr{\varepsilon_{\rm r}} \h \vec E\tot\,,
\end{equation}
which defines the microscopic dielectric tensor (see e.g.~\cite{Martin,SchafWegener}), the latter has to be characterized as the {\itshape total} functional derivative
\begin{equation}
\tsr{\varepsilon_{\rm r}} \h = \h \frac{\de\vec E\ext}{\de\vec E\tot} \h = \h \frac{\delta\vec E\ext}{\delta\vec E\tot}+\frac{\delta\vec E\ext}{\delta\vec B\tot} \h \frac{\delta\vec B\tot}{\delta\vec E\tot}\,.
\end{equation}
Here,  the derivative of the magnetic field with respect to the electric field has to be evaluated by means of Faraday's law (see \cite[\S\h4.2]{ED1}).
A similar equation holds for the physical conductivity tensor (see \cite[\S\h6.2]{ED1} or \cite[Eq.~(15)]{EDOhm}).
We particularly stress that the dielectric tensor must therefore not be identified with the partial response function given in Eq.~\eqref{eq_part_der1}
(see~\cite[Eq.~(12) and comments]{Melrose73} in this context).

We conclude that it is desirable to have relativistic transformation laws for the conductivity tensor and the dielectric tensor as well, since these are {\itshape physical}---and hence in principle measurable---response functions.
On the other hand, the {\it Cartesian} structure of the corresponding response relations casts serious doubts upon the shear existence of such relativistic transformation laws. For Ohm's law in the homogeneous limit, however, the proof of principle in this respect has been given by the authors of this article in Ref.~\cite{EDOhm}.
In the following, we will first rederive the already established \linebreak
transformation law of the conductivity tensor under Lorentz boosts by a method which displays a closer analogy to the original logic of H.~Minkowski's article \cite{Minkowski}. After that, we will deduce in a similar way the desired relativistic transformation law of the dielectric tensor.

\section{Cartesian boost transforms} \label{Sec_CovRT}

In this final section, we come to the problem of deriving a transformation law for the dielectric tensor under Lorentz boosts.
This problem has already been treated in Ref.~\cite{Melrose73} with a method which has later been rediscovered in Ref.~\cite{EDOhm}.
Unfortunately, however, Ref.~\cite{Melrose73} uses a relation between the dielectric tensor
and the conductivity tensor \cite[Eq.~(14)]{Melrose73} which is actually only valid in the long-wavelength limit (see \cite[\S\h2.5]{Refr}) and hence, in particular, not Lorentz covariant. In \S\h\ref{Subsec_LTsigma} and \S\h\ref{Subsec_LTeps}, we will therefore systematically derive the behavior of the conductivity tensor and the dielectric tensor
under Lorentz boosts. For technical reasons, however, we
will start this section with the derivation of transformation laws for the (spatial) current density (\S\h\ref{Subsec_LTj}) and the 
electric field vector in Fourier space (\S\h\ref{Subsec_LTE}).

\subsection{Spatial current density}\label{Subsec_LTj}

The Lorentz transformation law for the spatial current density can be derived by first transforming the four-current $j^\mu=(c\rho,\vec j)^{\rm T}$ in the unprimed reference frame with a Lorentz boost (see Appendix \ref{app_boost}) in order to obtain the four-current $j^{\mu'} = (c\rho', \vec j')^{\rm T}$ in the primed reference frame. In the resulting expression for $\vec j'$ in terms of $\vec j$ and $\rho$, one can then eliminate the charge density $\rho$ in terms of the current density $\vec j$ via the continuity equation.
The result will be a linear expression for the current density $\vec j'$ in the primed frame in terms of the current density $\vec j$ in the unprimed frame.

Concretely, our starting point is the fundamental transformation law for Minkowski tensor fields as applied to the current density four-vector in Fourier space, \smallskip
\begin{equation}
j^{\mu'}(k')=\Lambda\indices{^{\mu'}_\mu}\, j^\mu(k)\,, \label{eq_TrafoJ} \smallskip
\end{equation}
where $k' = \Lambda \h k$. The above equation is of course none other than Eq.~\eqref{usual_1} in Fourier space.
Using the expression \eqref{eq_boost} for the Lorentz boost $\Lambda = \Lambda(\vec v)$, we obtain from this the transformed current density in terms of the spatial part of the boost matrix as
\begin{equation} \label{eq_TrafoJ_zwischen}
\vec j'(\vec k', \omega') =-\gamma \h \rho(\vec k, \omega) \h \vec v+ \tsr\Lambda(\vec v) \, \vec j(\vec k, \omega)\,.
\end{equation}
On the other hand, by means of the continuity equation in Fourier space,
\begin{equation}
\rho(\vec k,\omega)=\frac{\vec k\cdot\vec j(\vec k,\omega)}{\omega}\,,
\end{equation}
we can eliminate the charge density from Eq.~\eqref{eq_TrafoJ_zwischen}. Thus, we find the relation for the spatial current density
\begin{equation}
\vec j'(\vec k',\omega')=\left(-\gamma \h \frac{\vec v\vec k^{\rm T}}{\omega}+\tsr\Lambda(\vec v) \right)\vec j(\vec k,\omega)\,,
\end{equation}
which can also be written as
\begin{equation}
\vec j'(\vec k',\omega')=\tsr\Lambda(\vec v)\left(\tsr 1-\gamma \, \tsr\Lambda{}^{-1}(\vec v) \h \frac{\vec v\vec k^{\rm T}}{\omega}\right)\vec j(\vec k,\omega)\,.
\end{equation}
Further using Eq.~\eqref{eq_invCartLor_corr}, we obtain
\begin{equation} \label{des_curr}
\vec j'(\vec k',\omega')=\tsr\Lambda(\vec v) \left(\tsr 1-\frac{\vec v\vec k^{\rm T}}{\omega}\right)\vec j(\vec k,\omega)\,.
\end{equation}
This is the desired relativistic transformation law for the spatial current density. For later purposes, we condense it into
\begin{equation}
\vec j'(\vec k',\omega')=\tsr\Lambda_{\vec j,\vec v}(\vec k,\omega) \, \vec j(\vec k,\omega)\,,
\end{equation}
where the transformation matrix is defined as 
\begin{equation}
\tsr\Lambda_{\vec j,\vec v}(\vec k,\omega):=\tsr\Lambda(\vec v) \left(\tsr 1-\frac{\vec v\vec k^{\rm T}}{\omega}\right) \,. \label{eq_TrafoMatrixJ}
\end{equation}
In particular, this shows that the transformation behavior of the spatial current density is not (and must not be) of the tensorial type, 
because the corresponding transformation matrix $\Lambda_{\vec j, \vec v}$ also depends on the arguments $(\vec k, \omega)$ of the vector field.
This is in contrast to the case of spatial rotations, which do not mix temporal and spatial components. In fact, under such rotations $R \in \mathrm{SO}(3)$, the current density transforms trivially as
\begin{equation}
 \vec j'(\vec k',\omega) =\tsr R\, \vec j(\vec k,\omega) \,,
\end{equation}
where $\vec k' = \tsr R \h \vec k$\hh. In other words, under spatial rotations the 
current density behaves as an ordinary Cartesian tensor. The same applies, of course, to the electric field vector, whose transformation under Lorentz boosts we will consider in the next subsection.

\subsection{Electric field vector}\label{Subsec_LTE}

In principle, the procedure used to derive the transformation law of the electric field is completely analogous to the one applied in the preceding subsection:
we start from the well-known behavior of the electric field under Lorentz boosts (see Appendix \ref{app_emtrafo}) and then express all quantities in terms of the electric field,
which is indeed feasible in Fourier space. However, in order to obtain a compact expression which suits our later purposes,
we will have to engage in a somewhat lenghty~calculation.

As a matter of principle, under a boost $\Lambda(\vec v)$ with velocity parameter $\vec v$, the electric field in Fourier space transforms according to Eq.~\eqref{eq_trafo1}, which we reproduce here again for the convenience of the reader:
\begin{equation} \label{eq_trafo_el}
\vec E'(\vec k',\omega')=\gamma \h \big(\vec E(\vec k,\omega)+\vec v\times\vec B(\vec k,\omega)\big)-(\gamma-1) \h \frac{\vec v\cdot\vec E(\vec k,\omega)}{|\vec v|^2} \, \vec v\,.
\end{equation}
At first sight, the electric field in the primed frame of reference also depends on the magnetic field in the unprimed frame.
In Fourier space, however, the magnetic field can be expressed in terms of the electric field as
\begin{equation}
\vec B(\vec k,\omega)=\frac{\vec k\times\vec E(\vec k,\omega)}{\omega}\,,
\end{equation}
which follows from Faraday's law (see \cite[\S\h4.2]{ED1}). Substituting this in the transformation law \eqref{eq_trafo_el} yields after a straightforward manipulation
\begin{equation}
\vec E'=\gamma \h \bigg(1-\frac{\vec v\cdot\vec k}{\omega}\bigg) \hh \vec E + (1-\gamma) \, \frac{\vec v\cdot\vec E}{|\vec v|^2} \, \vec v + \gamma \, \frac{\vec v\cdot\vec E}{\omega} \, \vec k\,,
\end{equation}
where the arguments of the electric field have been suppressed. By factoring out the coefficient of the electric field in the first term, this yields
\begin{equation}
\vec E'=\gamma \h \bigg(1-\frac{\vec v\cdot\vec k}{\omega}\bigg)
\bigg(\vec E + \frac{1-\gamma}{\gamma}\h \frac{\omega}{\omega-\vec v\cdot\vec k} \h 
\frac{\vec v \h (\vec v\cdot\vec E)}{|\vec v|^2}+\frac{\vec k \h (\vec v\cdot\vec E)}{\omega-\vec v\cdot\vec k} \hh\bigg)\,,
\end{equation}
and by factoring out the electric field itself on the right-hand side, we obtain
\begin{equation}
 \vec E'=\gamma \h \bigg(1-\frac{\vec v\cdot\vec k}{\omega}\bigg)\bigg(\hh\tsr 1+ \frac{1-\gamma}{\gamma} \h \frac{\omega}{\omega-\vec v\cdot\vec k} \h
\frac{\vec v\vec v^{\rm T}}{|\vec v|^2}+\frac{\vec k\vec v^{\rm T}}{\omega-\vec v\cdot\vec k}\hh\bigg) \hh \vec E\,.
\end{equation}
We now subject the second term in brackets to another complicated transformation. First, we re-express it in the form
\begin{equation}
\bigg(\hh \tsr 1+ 
\frac{1-\gamma}{\gamma} \h \frac{\vec v\vec v^{\rm T}}{|\vec v|^2}+
\frac{1-\gamma}{\gamma} \h \frac{\vec v\cdot\vec k}{\omega-\vec v\cdot\vec k} \h \frac{\vec v\vec v^{\rm T}}{|\vec v|^2}+
\frac{\vec k\vec v^{\rm T}}{\omega-\vec v\cdot\vec k}\hh\bigg) \hh \vec E\,.
\end{equation}
Next, using the projectors \eqref{def_pl} and \eqref{def_pt} on the direction of $\vec v$ and its complement, we recast this into
\begin{equation}
\bigg(\hh\frac{1}{\gamma} \h \tsr P\L(\vec v)+\tsr P\T(\vec v)+
\frac{1-\gamma}{\gamma} \, \tsr P\L(\vec v) \h \frac{\vec k\vec v^{\rm T}}{\omega-\vec v\cdot\vec k}+
\frac{\vec k\vec v^{\rm T}}{\omega-\vec v\cdot\vec k}\bigg) \hh \vec E\,.
\end{equation}
This expression can be further reworked to yield
\begin{equation}
\bigg(\hh\frac{1}{\gamma} \h \tsr P\L(\vec v) +\tsr P\T(\vec v) +
\frac{1}{\gamma} \h \tsr P\L(\vec v) \h \frac{\vec k\vec v^{\rm T}}{\omega-\vec v\cdot\vec k}+
\tsr P\T(\vec v) \h \frac{\vec k\vec v^{\rm T}}{\omega-\vec v\cdot\vec k}\hh\bigg) \hh \vec E\,.
\end{equation}

\pagebreak \noindent
Fortunately, the term in brackets now factorizes as
\begin{equation}
\bigg(\hh\frac{1}{\gamma} \h \tsr P\L(\vec v)+\tsr P\T(\vec v)\bigg)
\bigg(\hh\tsr1+\frac{\vec k\vec v^{\rm T}}{\omega-\vec v\cdot\vec k}\hh\bigg) \hh \vec E\,.
\end{equation}
Putting everything together, we obtain the transformation law for the electric field in the form
\begin{equation}
\vec E'=\gamma \h \bigg(1-\frac{\vec v\cdot\vec k}{\omega}\bigg)\bigg(\hh\frac{1}{\gamma} \h \tsr P\L(\vec v)+\tsr P\T(\vec v)\bigg)\bigg(\hh\tsr 1+\frac{\vec k \vec v^{\rm T}}{\omega-\vec v\cdot\vec k}\hh\bigg) \hh \vec E\,.
\end{equation}
Finally, by restoring arguments and using Eq.~\eqref{eq_invCartLor}, we arrive at
\begin{equation} \label{trafo_el_field}
\vec E'(\vec k',\omega')=
\gamma \h \bigg(1-\frac{\vec v\cdot\vec k}{\omega}\bigg) \, \tsr\Lambda{}^{-1}(\vec v) \, \bigg(\hh\tsr1+\frac{\vec k \vec v^{\rm T}}{\omega-\vec v\cdot\vec k}\hh\bigg) \hh \vec E(\vec k,\omega)\,.
\end{equation}
This is the desired transformation law for the electric field. In analogy to the 
case of the current density, we write it compactly as
\begin{equation} \label{trafo_el_field_compact}
\vec E'(\vec k',\omega')=\tsr\Lambda_{\vec E,\vec v}(\vec k,\omega) \h\hh \vec E(\vec k,\omega)\,,
\end{equation}
where the transformation matrix for the electric field is defined as
\begin{equation}
\tsr\Lambda_{\vec E,\vec v}(\vec k,\omega):=\gamma \h \bigg(1-\frac{\vec v\cdot\vec k}{\omega}\bigg) \, \tsr\Lambda{}^{-1}(\vec v) \,
\bigg(\hh\tsr 1-\frac{\vec k \vec v^{\rm T}}{\omega}\bigg)^{\!\!-1}\,. \label{eq_TrafoMatrixE}
\end{equation}
In this last formula we have further used the relation
\begin{equation}
\bigg(\hh\tsr 1-\frac{\vec k \vec v^{\rm T}}{\omega}\bigg)^{\!\!-1}=\bigg(\hh\tsr 1+\frac{\vec k \vec v^{\rm T}}{\omega-\vec v\cdot\vec k}\hh\bigg)\,,
\end{equation}
which is also straightforward to prove.

\subsection{Conductivity tensor}\label{Subsec_LTsigma}

In the preceding subsections, we have derived Cartesian transformation laws for the current density and for the electric field in Fourier space.
With these, we will now derive the resulting transformation law for the conductivity tensor. We start by assuming the validity of Ohm's law in the primed reference frame, i.e., \smallskip
\begin{equation} \label{usual_sig}
\vec j'= \tsr\sigma{}'\hh\vec E'\,. \smallskip
\end{equation}
Invoking the derived transformation laws in the symbolic form
\begin{align}
\vec j'&=\tsr\Lambda_{\vec j,\vec v}\,\h\vec j\,,\\[3pt]
\vec E'&=\tsr\Lambda_{\vec E,\vec v}\,\vec E\,,
\end{align}
and solving for the current density in the unprimed frame, we obtain
\begin{equation}
\vec j=\Big(\big(\hh\tsr\Lambda_{\vec j,\vec v}\big)^{\!-1} \,\h \tsr\sigma{}' \,\h \tsr\Lambda_{\vec E,\vec v}\Big) \hh \vec E\,.
\end{equation}
Interpreting this formula as Ohm's law in the unprimed frame---with the term in brackets being the conductivity tensor
in the unprimed frame---we find the transformation law for the conductivity tensor in the abstract form
\begin{equation}
\tsr\sigma{}'= \tsr\Lambda_{\vec j,\vec v} \,\hh \tsr\sigma \, \big(\hh\tsr\Lambda_{\vec E,\vec v}\big)^{\!-1}\,.
\end{equation}
After restoring all arguments by means of Eqs.~\eqref{eq_TrafoMatrixJ} and \eqref{eq_TrafoMatrixE}, this reads explicitly
\begin{align}\label{eq_trafosigma}
\tsr\sigma{}'(\vec k',\omega') & = \frac{1}{\gamma}\left(1-\frac{\vec v\cdot\vec k}{\omega}\right)^{\!\!-1} \,\tsr\Lambda(\vec v)\, \bigg(\hh\tsr 1-\frac{\vec v\vec k^{\rm T}}{\omega}\bigg) \, \tsr\sigma(\vec k,\omega) \, \bigg(\hh\tsr 1-\frac{\vec k \vec v^{\rm T}}{\omega}\bigg) \, \tsr\Lambda(\vec v)\,.
\end{align}
This is the searched-for transformation law for the conductivity tensor. It coincides exactly with \cite[Eq.~(28)]{EDOhm}, where it was derived 
by a completely different method (whose logic, however, can already be found in the seminal work Ref.~\cite{Melrose73}). In the same article, 
we have shown that the assumption of a scalar and constant conductivity in the primed reference frame
yields back the ordinary relativistic generalization of Ohm's law, Eq.~\eqref{eq_OhmGen}.

\subsection{Dielectric tensor}\label{Subsec_LTeps}

Finally, we come to the Lorentz boost transform of the dielectric tensor. We proceed in complete analogy to the case of the conductivity tensor: we start from the defining equation for the dielectric tensor in the primed frame of reference, \smallskip
\begin{equation} \label{usual_eps}
\vec E'\ext= \tsr{\varepsilon_{\rm r}}{}' \h \vec E'\tot\,, \smallskip
\end{equation}
and substitute Eq.~\eqref{trafo_el_field_compact} for both the external and the total electric field. In this way we find the fundamental transformation law of the dielectric tensor \linebreak

\pagebreak \noindent
in the symbolic form \smallskip
\begin{equation}
\tsr{\varepsilon_{\rm r}}{}'=\tsr\Lambda_{\vec E,\vec v} \,\h  \tsr{\varepsilon_{\rm r}} \,\hh \big(\hh\tsr\Lambda_{\vec E,\vec v}\big)^{\!-1}\,. \smallskip
\end{equation}
Using the concrete expression \eqref{eq_TrafoMatrixE}, we  arrive at the compact formula
\begin{equation}
\tsr{\varepsilon_{\rm r}}{}'(\vec k',\omega')= \tsr\Lambda{}^{-1}(\vec v) \, \bigg( \hh \tsr 1-\frac{\vec k\vec v^{\rm T}}{\omega} \bigg)^{\!\!-1}  
\tsr{\varepsilon_{\rm r}}(\vec k,\omega) \, \bigg( \hh \tsr 1-\frac{\vec k\vec v^{\rm T}}{\omega}\bigg) \, \tsr\Lambda(\vec v)\,. \label{eq_trafoeps}
\end{equation}
This is the transformation law for the wavevector- and frequency-dependent dielectric tensor, which constitutes the central result of this article.

Apart from the general interest which the correct relativistic transformation
law of the dielectric tensor certainly deserves, there are essentially three important conclusions to be drawn from our result~\eqref{eq_trafoeps}:

\begin{enumerate}[listparindent=\parindent,parsep=0pt,itemsep=1em]
 \item[(i)]{\itshape Scalar dielectric constant.}
First, it follows that the assumption of a constant dielectric function (i.e., a dielectric function not depending on $\omega$ and $\vec k$) is Lorentz
covariant. Thus, if the dielectric tensor is given by a {\itshape scalar constant} in the unprimed inertial frame,
\begin{equation}
 (\varepsilon_{\rm r})_{ij}(\vec k, \omega) = \varepsilon_{\rm r} \, \delta_{ij} \,,
\end{equation}
then it is also given by a scalar constant in the primed inertial frame,
\begin{equation}
 (\varepsilon'_{\rm r})_{ij}(\vec k', \omega') = \varepsilon'_{\rm r} \, \delta_{ij} \,,
\end{equation}
and moreover, these constants coincide,
\begin{equation}
 \varepsilon_{\rm r}' = \varepsilon_{\rm r} \,.
\end{equation}
 Put differently, {\it the dielectric constant is a Lorentz scalar}.
This is in stark contrast to the conductivity tensor, which can be assumed constant at most in a single inertial frame. In all boosted inertial frames, it will then have a wavevector and frequency dependence which is determined by the corresponding transformation law \eqref{eq_trafosigma}.

 \item[(i)]{\itshape Scalar dielectric function.}
Second, if the dielectric tensor is given by a dielectric function times the unit matrix, i.e.,
\begin{equation}
\tsr{\varepsilon_{\rm r}}(\vec k,\omega)=\varepsilon_{\rm r}(\vec k,\omega) \h \tsr 1\,,
\end{equation}
then this structure is also preserved under Lorentz transformations, and the involved dielectric function transforms trivially as
\begin{equation}
\varepsilon'_{\rm r}(\vec k',\omega')=\varepsilon_{\rm r}(\vec k,\omega)\,.
\end{equation}
In other words, the dielectric function transforms as a scalar function under Lorentz boosts. 

\item[(iii)]{\itshape Dispersion relation.}
Finally, by taking the determinant of the transformation law \eqref{eq_trafoeps}, we obtain the equality
\begin{equation}\label{eq_EqDetDieTens}
\det\h\hh\tsr{\varepsilon_{\rm r}}{}'(\vec k',\omega')=\det\h\hh\tsr{\varepsilon_{\rm r}}(\vec k,\omega)\,.
\end{equation}
This is important for the following reason: The electromagnetic dispersion relation in materials, $\omega=\omega_{\vec k}$, is defined by 
the condition (see Ref.~\cite{Refr} for a paradigmatic discussion)
\begin{equation}
\det\h\hh\tsr{\varepsilon_{\rm r}}(\vec k,\omega_{\vec k})=0\,. \label{eq_DefDispMedia}
\end{equation}
From the equality \eqref{eq_EqDetDieTens} it now follows that the four-vector $(\omega_{\vec k}/c, \h \vec k)^{\rm T}$, 
which describes proper oscillations of the medium in the unprimed inertial frame, corresponds to the four-vector $(\omega'_{\vec k'}/c, \h \vec k')^{\rm T}$
in the primed inertial frame. In other words,
the dispersion relation $\omega_{\vec k}$ can be combined with the wavevector $\vec k$ into a four-vector, and this {\it formal} four-vector transforms precisely as it should:
namely as a {\it Minkowski} four-vector. We illustrate this result in Appendix~\ref{sec_fizeau} by applying 
it to the transformation behavior of the refractive index.
\end{enumerate}

\smallskip \noindent
We conclude this section with the following remark: The conductivity tensor and the dielectric tensor each contain separately the full information about the linear electromagnetic response of any material. The reason for this is that any magnetic field can be expressed via Faraday's law in terms of a transverse electric field. On the other hand, the longitudinal part of the electric field can in general not be eliminated in terms of the magnetic field, and consequently, the magnetic permeability tensor or the magnetoelectric cross-coupling tensors do not contain the complete information about the linear electromagnetic response (see Ref.~\cite{ED1}). Therefore, it is also not possible to formulate a relativistic transformation law for these latter response functions. Furthermore, as we have pointed out already in Ref.~\cite{EDOhm}, a description in terms of the  conductivity tensor or the dielectric tensor alone is also not possible in the presence of {\itshape static} magnetic fields (which we have implicitly excluded in this article). This restriction, however, does not concern the relativistic transformation laws of the corresponding material relations but rather their very validity in {\itshape any} inertial frame.

\section{Conclusion}

After a short critique of the Minkowski formulae, we have given a concise introduction to modern microscopic electromagnetic response theory as
embodied in particular in the Functional Approach to electrodynamics of media. Concomitantly, we have shown that this approach is inherently Lorentz covariant
and thus not in need of a relativistic generalization. On the most fundamental level, a Lorentz-covariant description of electromagnetic material properties can be given in terms of the fundamental response tensor (\mbox{\S\h\ref{sec_fundresp}}) or in terms of the field strength response tensor (\S\h\ref{sec_fsrt}). On the other hand, we have shown in this article that a {\itshape covariant Cartesian response theory} in terms of the  conductivity tensor or the dielectric tensor can be formulated as well. The corresponding response laws, Eq.~\eqref{usual_sig} and Eq.~\eqref{usual_eps}, are in fact already Lorentz covariant, provided that {\itshape all} quantities involved in the response laws (the fields {\itshape and} the response functions) are correctly transformed under coordinate changes.
The transformation laws under Lorentz boosts are given explicitly for the spatial current density by Eq.~\eqref{des_curr}, for the electric field by Eq.~\eqref{trafo_el_field}, for the conductivity tensor by Eq.~\eqref{eq_trafosigma}, and for the dielectric tensor by Eq.~\eqref{eq_trafoeps}. Our derivation of these fundamental Lorentz transformation laws follows a logic closely parallel to the original derivation of the Minkowski formulae. Finally, we have discussed the special cases of a scalar dielectric function and a dielectric constant, and we have justified the transformation law of the dispersion relation, which has already been used in Ref.~\cite[Appendix A]{Refr} for investigating the refractive index of moving media. Our results show that modern {\itshape ab initio} materials physics also leads to conceptually new results in classical relativistic electrodynamics, an area of theoretical physics which appeared to be definitely settled for more than a century.

\pagebreak \noindent
\section*{Acknowledgments}

This research was supported by the DFG grant HO 2422/12-1 and by the DFG RTG 1995. R.\,S. thanks the Institute for Theoretical Physics at TU Bergakademie Freiberg for its hospitality.

\begin{appendices}

\section{Basics of special relativity} \label{app_srt}

\subsection{Lorentz boost} \label{app_boost}

In this appendix, we review some aspects of special relativity relevant to this article.
For more detailed introductions, see e.g.~Refs.~\cite{Fliessbach,Scheck,Sexl,Misner}.

Generally, the Lorentz group consists of linear transformations $\mathbb R^4\rightarrow\mathbb R^4$,
\begin{equation} \label{eq_sugg}
x^{\mu'} = \Lambda\indices{^{\mu'}_\mu} \h x^\mu  \,,
\end{equation}
which leave invariant the Minkowski scalar product
\begin{equation}
 \eta(x,y) \h \equiv \h \eta_{\mu\nu} \h x^\mu \hh y^\nu \h = \h -x^0y^0 +\vec x\cdot\vec y \,.
\end{equation}
Here, we have used a suggestive index notation which emphasizes the facts that (i) the vector $x$ on both sides of Eq.~\eqref{eq_sugg} is the same vector but expressed in different (primed or unprimed) coordinates, and (ii) the transformation matrix $\Lambda\indices{^{\mu'}_\mu}$ is itself not a Lorentz tensor, as its two indices $\mu'$ and $\mu$ refer to different coordinate systems. Physically, a Lorentz transformation corresponds to a coordinate change between two inertial systems. This means, the space-time coordinates $x^\mu$ and $x^{\mu'}$
describe the same event but in different inertial systems: \smallskip
\begin{equation}
 x^\mu \vec e_\mu = x^{\mu'}\vec e_{\mu'} \,, \smallskip
\end{equation}
where $\vec e_\mu$ and $\vec e_{\mu'}$ are the respective orthonormal basis vectors of the two coordinate systems (pointing in the respective time and space directions).
One easily convinces oneself that the Lorentz group comprises the time-reversal transformation, \smallskip
\begin{equation}
 x^0\mapsto -x^0 \,, \smallskip
\end{equation}
the spatial reflection (parity transformation),
\begin{equation}
 \vec x\mapsto-\vec x \,,
\end{equation}

\pagebreak \noindent
and all spatial rotations, \smallskip
\begin{equation}
 \vec x\mapsto \tsr R\, \vec x\,, \smallskip
\end{equation}
where $R\in{\rm SO}(3)$. While spatial rotations, time-reversal and spatial reflection are also symmetries in Galilean mechanics, the qualitatively
new effects of special relativity are embodied in the {\itshape Lorentz boosts,} which are the relativistic analogon of the {\itshape special Galilei transformations} 
(see e.g.~\cite[\S\h4.4.2]{Scheck}).
Fortunately, every ``proper, orthochronous'' Lorentz transformation can be represented as the product of a spatial rotation and a Lorentz boost, 
while the most general Lorentz transformation is a ``proper, orthochronous'' Lorentz transformation possibly combined with the time-reversal and/or the parity transformation. For many purposes, it therefore suffices to study only the Lorentz boosts.

Concretely, a Lorentz boost matrix is of the form
\begin{equation}\label{eq_boost}
\Lambda\indices{^\mu_\nu}(\vec v)=\left( \!
\begin{array}{cc}
\gamma & -\gamma\vec v^{\rm T}/c\\[5pt]
-\gamma\vec v/c & \tsr \Lambda(\vec v)
\end{array} \!
\right),
\end{equation}
with the {\itshape relativistic root}
\begin{equation}
\gamma=\frac{1}{\sqrt{1-|\vec v|^2/c^2}} \,,
\end{equation}
and the spatial part of the boost matrix given by
\begin{equation}
\tsr\Lambda(\vec v) = \tsr{1}+(\gamma-1) \h \frac{\vec v \hh \vec v^{\rm T}}{|\vec v|^2} \,. \label{eq_spatialBoost}
\end{equation}
We note that both this $(3 \times 3)$-matrix and the whole $(4\times 4)$-matrix in Eq. \eqref{eq_boost} are Hermitean, i.e., \smallskip
\begin{equation}
 \Lambda(\vec v) = \Lambda^{\rm T}(\vec v) \,. \smallskip
\end{equation}
For $\Lambda = \Lambda(\vec v)$, the transformation \eqref{eq_sugg} of the time and space coordinates can be written explicitly as
\begin{align}
 t' & = \gamma \h \bigg( t - \frac{\vec v \cdot \vec x}{c^2} \bigg) \,, \\[3pt]
 \vec x' & = \vec x + \vec v \h \bigg( {-\gamma \hh t + (\gamma - 1) \h \frac{\vec v \cdot \vec x}{|\vec v|^2}} \bigg) \,.
\end{align}
Physically, the vector-valued boost parameter $\vec v$ is therefore the velocity of the primed coordinate frame relative to the unprimed frame. (This becomes immediately clear by considering the origin of the primed frame, which has the spatial coordinates $\vec x' = 0$ and $\vec x = \vec v \hh t$, respectively.)
Correspondingly, the inverse of~$\Lambda(\vec v)$ is given by
\begin{equation} \label{no_conflict}
 \Lambda^{-1}(\vec v) = \Lambda(-\vec v) \,.
\end{equation}
Finally, we assemble some useful formulae. First, the relativistic root fulfills the identity \smallskip
\begin{equation} \label{eq_relroot}
\gamma^2=\frac{c^2}{|\vec v|^2} \h (\gamma+1)(\gamma-1)\,. \smallskip
\end{equation}
On the other hand, we can write the spatial part of the Lorentz boost, Eq. \eqref{eq_spatialBoost}, in the alternative form
\begin{equation}\label{eq_CartLor}
\tsr \Lambda(\vec v) \h =  \h \tsr 1 + (\gamma - 1) \h \tsr P\L(\vec v) \h = \h \gamma \h \tsr P\L(\vec v)+\tsr P\T(\vec v)\,,
\end{equation}
with the longitudinal and transverse (with respect to the velocity $\vec v$) projection operators
\begin{align}
\tsr P\L(\vec v)&=\frac{\vec v \hh \vec v^{\rm T}}{|\vec v|^2}\,, \label{def_pl} \\[5pt]
\tsr P\T(\vec v)&=\tsr 1-\tsr P\L(\vec v) \,. \label{def_pt}
\end{align}
For the inverse of \eqref{eq_CartLor}, we then have
\begin{equation}\label{eq_invCartLor}
\big( \hh \tsr\Lambda(\vec v)\big)^{-1} \equiv \h \tsr \Lambda{}^{-1}(\vec v)  \h = \h \frac{1}{\gamma} \h \tsr P\L(\vec v)+\tsr P\T(\vec v)\,,
\end{equation}
whence we get the relation
\begin{equation} \label{eq_invCartLor_corr}
\tsr\Lambda{}^{-1}(\vec v) \, \vec v=\frac{\vec v}{\gamma}\,.
\end{equation}
We particularly emphasize that for the Cartesian boost matrix \eqref{eq_invCartLor},
\begin{equation}
 \tsr \Lambda{}^{-1}(\vec v) \neq \tsr \Lambda(-\vec v)\,,
\end{equation}
which is not in conflict with the property \eqref{no_conflict} of the four-dimensional (Minkowskian) Lorentz boost matrix.

\subsection{Electromagnetic field transformation} \label{app_emtrafo}

In this subsection, we assemble the relevant formulae for the Lorentz boost transformations of electromagnetic fields. 
We give a derivation in the $(3+1)$-formalism (see e.g.~Refs.~\cite{Alcubierre,GambiniPullin}, and for another recent application see Ref.~\cite{EDFullGF}). First, we recall that the field strength tensor is defined as
\begin{equation}
F^{\mu\nu}=\partial^\mu A^\nu - \partial^\nu A^\mu\,.
\end{equation}
With the definition of the four-potential $A^\mu$ in terms of the scalar potential $\varphi$ and the vector potential $\vec A$, i.e.,
\begin{equation}
A^\mu=\bigg( \!\! \begin{array}{c} \varphi/c \\[3pt] \vec A \end{array} \!\mh \bigg) \,,
\end{equation}
this yields the entries of the field strength tensor as
\begin{align}
F^{0i}&=E^i/c\,,\\[3pt]
F^{ij}&=\epsilon_{ijk}B_k\,.
\end{align}
We further introduce the Cartesian {\itshape magnetic field matrix}
\begin{equation} \label{def_Bmat}
B_{ij} := \epsilon_{ikj}B_k\,,
\end{equation}
from which the magnetic field vector can be retrieved as
\begin{equation} \label{Bvecmat}
 B_i = \frac 1 2 \h \epsilon_{jik} B_{jk} \,.
\end{equation}
In matrix form, we can write Eq.~\eqref{def_Bmat} as
\begin{equation}
\tsr{\vec B}=
\left(
\begin{array}{ccc}
0 & -B_3 & B_2\\[1pt]
B_3 & 0 & -B_1\\[1pt]
-B_2 & B_1 & 0
\end{array}
\right).
\end{equation}
In particular, the magnetic field matrix is antisymmetric and has the following properties for every vector $\vec v$:
\begin{align}
\tsr{\vec B}\vec v&=\vec B\times\vec v\,, \label{eq_Bv} \\[3pt]
\vec v^{\rm T}\tsr{\vec B}&=(\vec v\times\vec B)^{\rm T}\,. \label{eq_vTB}
\end{align}
The field strength tensor can now be rewritten in the $(3+1)$-formalism as
\begin{equation} \label{unpr_fst}
F^{\mu\nu}=\left( \!
\begin{array}{cc}
0 & \vec E^{\rm T}/c \\[5pt]
-\vec E/c &-\tsr{\vec B}
\end{array} \!
\right).
\end{equation}
We now come to the behavior of the electric and magnetic fields under Lorentz transformations, in particular under Lorentz boosts. As a matter of principle, the field strength tensor transforms as a second-rank tensor, i.e.,
\begin{equation}
F^{\mu'\nu'}=\Lambda\indices{^{\mu'}_\mu} \h \Lambda\indices{^{\nu'}_\nu} \h F^{\mu\nu}\,.
\end{equation}
This can be written equivalently as a matrix multiplication,
\begin{equation}
F'=\Lambda \h  F \h \Lambda^{\rm T}\,, \label{eq_TrafoF}
\end{equation}
with the boost matrix $\Lambda \equiv \Lambda(\vec v)$ given by Eq.~\eqref{eq_boost}. This transformation law preserves the antisymmetry of the field strength tensor, $F^{\rm T}=-F$,
and hence we can make the ansatz for the transformed field strength tensor
\begin{equation} \label{primed_ansatz}
F^{\mu'\nu'}=\left( \!
\begin{array}{cc}
0 & \vec E'^{\hh\rm T}\mh/c \\[5pt]
-\vec E'\mh/c &-\tsr{\vec B}{}'
\end{array} \!
\right),
\end{equation}
which implicitly defines the electric and magnetic fields in the primed coordinate frame. Now, using Eqs.~\eqref{eq_boost} and \eqref{unpr_fst}, we can multiply out the right-hand side of Eq.~\eqref{eq_TrafoF} in the $(3+1)$-formalism and compare the result to Eq.~\eqref{primed_ansatz}. With the help of the identities \eqref{eq_relroot} and \eqref{eq_Bv}--\eqref{eq_vTB}, we then find the transformation law for the electric field vector,
\begin{equation}
\vec E'	=	-\frac{\gamma^2}{c^2}(\vec v\cdot\vec E) \h \vec v + \gamma \h \tsr\Lambda \h \vec E - \gamma \h \tsr\Lambda \h \tsr{\vec B} \h \vec v\,,
\end{equation}
which after some manipulations simplifies to
\begin{equation}
\vec E'=\gamma\h(\vec E+\vec v\times\vec B)-(\gamma - 1) \h \frac{\vec v \h (\vec v\cdot\vec E)}{|\vec v|^2}\,. \label{eq_trafo1} 
\end{equation}
Similarly, for the magnetic field matrix, we first find
\begin{equation}
 \tsr{\vec B}{}'= \frac{\gamma}{c^2} \h \vec v \h (\vec E^{\rm T}\tsr\Lambda) - \frac{\gamma}{c^2} \h (\tsr\Lambda \h \vec E) \h \vec v^{\rm T} + \tsr\Lambda \h \tsr{\vec B} \h \tsr\Lambda\,,
\end{equation}
which can be rewritten as
\begin{equation} \label{zwischen_1}
 \tsr{\vec B}{}' = \tsr{\vec B} + \vec w \vec v^{\rm T} - \vec v \vec w^{\rm T} \,,
\end{equation}
where we have defined the vector
\begin{equation} \label{zwischen_2}
 \vec w = -\gamma \h \frac{\vec E}{c^2} + (\gamma - 1) \h \frac{\vec B \times \vec v}{|\vec v|^2} \,.
\end{equation}
In components, Eq.~\eqref{zwischen_1} means
\begin{equation}
 B'_{jk} = B_{jk} + w_j v_k - v_j w_k \,.
\end{equation}
Using Eq.~\eqref{Bvecmat}, this implies for the magnetic field components that
\begin{equation}
 B'_i = B_i - \epsilon_{ijk} \h w_j \h v_k \,,
\end{equation}
or equivalently, in vector notation,
\begin{equation}
 \vec B' = \vec B - \vec w \times \vec v \,.
\end{equation}
Substituting Eq.~\eqref{zwischen_2} for $\vec w$ therefore yields
\begin{equation}
\vec B'=\gamma\left(\vec B-\frac{\vec v\times\vec E}{c^2}\right)- (\gamma - 1) \, \frac{\vec v \h (\vec v \cdot \vec B)}{|\vec v|^2} \,, \label{eq_trafo2} 
\end{equation}
which is the desired transformation law for the magnetic field. In particular, the results \eqref{eq_trafo1} and \eqref{eq_trafo2} imply that
\begin{align}
\vec v \cdot \vec E' & = \vec v \cdot \vec E \,, \label{eq_boost_parr_1} \\[3pt]
\vec v \cdot \vec B' & = \vec v \cdot \vec B\,, \label{eq_boost_parr_2}
\end{align}
which means that the electric and magnetic field components parallel to the boost velocity $\vec v$ stay constant under the Lorentz boost transformation.

\section{Fizeau experiment and Fresnel drag} \label{sec_fizeau}

To illustrate the covariant definition of the dispersion relation in materials, we here re-investigate theoretically
the Fizeau experiment. This experiment established that the speed of light $u'$ in a moving medium is related to the speed $u$
in the respective medium at rest by the conspicuous formula \cite[pp.~447~f.]{Joos}
\begin{equation}
u'=u-v\left(1-\frac{1}{n^2}\right).
\end{equation}
Here, $v$ is the speed of the medium, while the term in brackets is known as the {\it Fresnel drag coefficient} \cite[p.~1154]{Bergmann}.
This result was later spectacularly explained by M.~von Laue, 
who deduced it from the relativistic velocity addition theorem,
\begin{equation} \label{eq_laue}
u'=\frac{u-v}{1-u \h v/c^2} \,, \smallskip
\end{equation}
by means of a Taylor expansion \cite{Laue}.

Before showing that our approach is consistent with these results,
let us first note some problems of the na\"{i}ve employment of Eq.~\eqref{eq_laue}.
In fact, it is a~priori not clear whether the ``speed of light in media'' can be used as a bona fide velocity to be plugged
into the relativistic velocity addition formula as if it would correspond to the velocity of a material particle.
Concretely, when the refractive index becomes frequency dependent, $n=n(\omega)$, so does the speed of light $u$ in the medium.
Consequently, the question arises to which frequency the transformed speed $u'$ belongs. In particular,
there is no transformed frequency $\omega'$ determined by the original frequency $\omega$ alone, because
any frequency is part of a four-vector $(\omega/c,\, \vec k)$ and hence transforms according to
\,$\omega'=\gamma\h (\omega-\vec v \cdot \vec k)$
\h under a Lorentz boost with the velocity $\vec v$. The transformed frequency is therefore undetermined if no wave-vector is specified. 

\begin{figure}[t]
\begin{center}
\includegraphics[width=\textwidth]{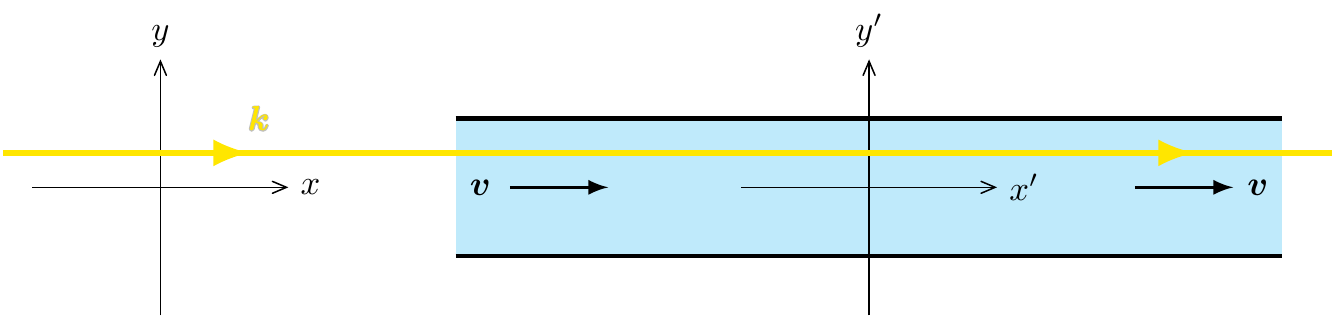}
\caption{Schematic representation of the Fizeau experiment. The unprimed coordinate system corresponds to the laboratory frame, while the primed coordinate system moves with the velocity $\vec v$ of the medium relative to the laboratory frame. In the case considered here, the wave-vector $\vec k$ of the electromagnetic wave is parallel to the velocity~$\vec v$. \label{fig_fizeau}} \vspace{0.3cm}
\end{center}
\end{figure}

In order to resolve these problems, we recall that in our approach the speed of light is defined by the dispersion relation, see Eq.~\eqref{eq_DefDispMedia}. 
The dispersion relation in turn defines the four-vector
\begin{equation}
k^\mu \h \equiv \h (\omega_{\vec k}/c,\,\vec k)^{\rm T} = (u_{\vec k}\hh|\vec k|\hh/\hh c,\,\vec k)^{\rm T}\,,
\end{equation}
from which the speed of light can be gained back as (see Ref.~\cite{Refr} for an extensive discussion)
\begin{equation}
u_{\vec k} =\frac{\omega_{\vec k}}{|\vec k|} = \frac{ck^0}{|\vec k|} \,. \smallskip
\end{equation}
Within our first-principles approach, it is therefore mandatory  to consider the whole four-vector $k$, for which the Lorentz transformation is of course well-defined. We will now show
that this procedure indeed reproduces M.\,\h von Laue's result. For simplicity, we consider a Lorentz boost in the direction of the 
wave-vector $\vec k$ (see the schematic representation in Fig.~\ref{fig_fizeau}). The wave-vector and the frequency then transform as
\begin{align}
\omega'&=\gamma \h (\omega-v\hh |\vec k|) \,, \\[5pt]
\vec k'&=\gamma \h (|\vec k|-\omega \h\hh v/c^2)\h\hh \vec e_{\vec k}\,,\label{eq_k-trafo}
\end{align}
where $\gamma = (1 - v^2/c^2)^{-1/2}$, and where $\vec e_{\vec k}$ is the unit vector in the direction of~$\vec k$. For the transformed speed of light in the medium, we now find
\begin{equation} \label{res}
u'_{\vec k'} =\frac{\omega'_{\vec k'}}{|\vec k'|}=\frac{\omega_{\vec k} - v\hh|\vec k|}{|\vec k|-\omega_{\vec k} \hh v/c^2}=\frac{u_{\vec k}-v}{1-u_{\vec k} \hh v/c^2}\,.
\end{equation}
This is precisely the desired relativistic velocity addition law. Our considerations show that, as a matter of principle,
one always has to consider the whole four-vector $(\omega/c, \h \vec k)$ together with the dispersion relation $\omega=\omega_{\vec k}$ of the material. For example, one can then also consider the case where the wave-vector and the boost velocity are not parallel.
Finally, we remark that already H.\,\h Weyl deduced the Fizeau result from the relativistic transformation properties of the four-vector $(\omega_{\vec k}/c, \h\vec k)$ (see \cite[p.~156]{Weyl}).

\end{appendices}

\bibliographystyle{model1-num-names}
\bibliography{/net/home/lxtsfs1/tpc/schober/Ronald/masterbib}

\end{document}